\documentstyle[12pt]{article}

\begin{document}

\begin{titlepage}
\begin{flushright}
PUPT-2010    \\

hep-th/0110196
\end{flushright}

\vspace{7 mm}

\begin{center}
{\huge Gauge Fields and Space-Time}
\end{center}
\vspace{10 mm}
\begin{center}
{\large

A.M.~Polyakov\\
}
\vspace{3mm}
Joseph Henry Laboratories\\
Princeton University\\
Princeton, New Jersey 08544
\end{center}
\vspace{7mm}
\begin{center}
{\large Abstract}
\end{center}
\noindent
In this article I attempt to collect some  ideas,opinions and formulae
which may be useful in solving the problem of gauge/ string / space-time correspondence
This includes the validity of D-brane representation, counting of gauge-invariant words,
relations between the null states and the Yang-Mills equations and the discussion of the strong
coupling limit of the string sigma model. The article is based on the talk given at the "Odyssey
2001" conference.
\vspace{7mm}
\begin{flushleft}
October 2001

\end{flushleft}
\end{titlepage}

\section{Introduction}

As time goes by, it seems that in fundamental physics we are circling around
a particular set of themes and ideas. One of the important examples of such
''limit cycles'' is the relations between gauge fields and strings. The
logic of these connections is as following. We begin with an observation (by
K. Wilson [1] ) that in the strong coupling limit of a lattice gauge theory
the elementary excitations are represented by closed strings formed by the
color-electric fluxes. In the presence of quarks these strings open up and
end on the quarks, thus guaranteeing quark confinement. More over, in the
SU(N) gauge theory the strings interaction is weak at large N. This fact
makes it reasonable to expect that also in the physically interesting
continuous limit ( not accessible by the strong coupling approximation) the
best description of the theory should involve the flux lines (strings) and
not fields, thus returning us from Maxwell to Faraday. In other words it is
natural to expect an exact duality between gauge fields and strings. The
challenge is to build a precise theory on the string side of this duality.

It is also possible to turn this problem around. We know that there is a
class of string theories, the superstrings, which contain quantum gravity
and ( as the enthusiasts say) everything else. It might be possible to
visualize the superstrings as flux lines of some unknown gauge theory. That
would give us a completely novel view of what we call a space -time. The
view is that in a deep sense it does not exist, being only a quasiclassical
limit of some abstract gauge theory, residing nowhere. Instead, our
observables must be sets of gauge invariant operators, formed out of the
products of some elementary ones, just as words are formed from the
alphabet. The theory should provide us with the expectation values for the
different combination of letters.

In recent years a serious progress has been achieved in realization of these
ideas ( for a review see [2] and [3]). In this lecture I will discuss some
of my latest attempts to find the dynamical correspondence between strings
and words. The main idea of the discussion below is quite simple. String
theories possess infinite number of gauge symmetries in the target space,
generated by the sequence of the zero norm states on the world sheet. The
lowest of these symmetries is general covariance. It is well known that the
general covariance on the string side leads to the conservation of the
energy momentum tensor of the gauge theory. We will see that the higher
gauge symmetries generate relations between the words, satisfied as a
consequence of the equations of motion of the gauge theory.

\section{ The general picture}

The origin of the gauge fields in general / strings duality (and AdS/CFT in
particular) can be traced to the relation between open and closed strings,
known from the very beginning of string theory. This relation in its
simplest form tells us the amplitude of an annulus represents either a
partition function of the open string or a propagation of the closed one. In
the d+1-dimensional space the amplitude the annulus with the external
particles attached to its outer circle and the inner circle placed at the
distance $R$, has a representation
\begin{equation}
Z=\int_{0}^{\infty }\frac{d\tau }{\tau ^{\frac{d}{2}}}f(\tau )\exp (-\frac{%
R^{2}}{2\tau })\int_{0}^{2\pi }\prod d\alpha _{k}\exp (-\sum
p_{i}p_{j}\Delta (\alpha _{i}-\alpha _{j}\mid \tau ))
\end{equation}
where we assumed the Dirichlet boundary condition in the d+1 direction,
Neuman condition in all the others and take $p_{i}$ to be d -dimensional
momenta ; $f(\tau )$is the corresponding determinant and $\Delta $ is a
propagator on a cylinder
\begin{equation}
\Delta (\alpha )=\sum_{n\neq 0}\frac{e^{in\alpha }}{n\tanh (n\tau )}
\end{equation}
We see that for the small $\tau $ (a thin cylinder) this propagator is that
of a particle on a circle, while as $\tau \rightarrow \infty $ the inner
circle of the annulus can be replaced by an insertion of a local operator.
In the leading order in the bosonic string this operator is the closed
string tachyon. For a general value of $\tau $ it is still possible to
replace the inner circle by a combination of local operators $\{O_{n}\}.$%
These operators are the Ishibashi states, annihilated by the boundary energy
momentum tensor. They are related to the primary operators $V_{n}$ by the
well known formulas
\begin{equation}
O_{n}=(1+\frac{1}{2\Delta }L_{-1}\overline{L_{-1}}+...)V_{n}
\end{equation}
where $\Delta $ is the dimension of $V.$ The contribution of an annulus is
equal to the contribution of a disc with an operator $V_{H}$ , describing
the hole, inserted at the middle. This operator has an expansion
\begin{equation}
V_{H}=\sum_{n}c_{n}(\tau )O_{n}
\end{equation}
with the coefficient functions behaving as
\begin{equation}
c_{n}\sim e^{-n\tau }
\end{equation}
at large $\tau .$ When we have several holes in the world sheet they all can
be replaced by the hole operators $V_{H}$.

In order to get a general idea of gauge/ strings correspondence, imagine for
a moment that this procedure converges ( which is not true in a simple
bosonic string). We know that we can obtain the Yang -Mills amplitudes from
the string by taking a limit of zero slope. In this limit the integral in
(1) is dominated by the small $\tau $ and represents a loop of a vector
particle. When we have a disk with several holes, it will shrink to a planar
Feynman diagram of the Yang -Mills theory. But if there is a convergence in
(1), the same set of diagrams can be viewed as a string disk diagram in some
general background generated by the $V_{H}$ insertions. We would conclude
that the amplitudes of gauge theory in d dimensions are equal to the
amplitudes of string theory in some non-trivial background field.

However there is a possible divergence of the above procedure which can
easily invalidate it. Namely, if the theory possesses an open string
tachyon, it will manifest itself in the divergence at small $\tau .$ Hence
we should be looking for theories without such tachyons. As was suggested in
[2] from a different point of view, we must consider a non-critical
fermionic NSR string with the non-chiral GSO projection . The gravitational
background (induced by the hole operators) for these strings must have the
form
\begin{equation}
ds^{2}=d\varphi ^{2}+a^{2}(\varphi )dx^{2}
\end{equation}
where the $\varphi $ direction comes from the Liouville field of the
non-critical string theory. The GSO projection comes from the summation on
the spin structures of the NSR fermions. This summation is indispensable for
having modular invariant amplitudes. From the open string point of view it
eliminates the open string tachyon. From the point of view of the closed
string background, allowing the world sheet fermions to change sign as we go
around the hole, means that there must be the Ramond- Ramond fields in the
background, coupled to the spin operators on the world sheet. Thus we have
the connection: {\it modular invariance - absence of the open tachyons - RR
background. }

If the tachyon is absent for other reasons this is just as well. In this
case the RR background is not needed. As an example consider c$\leq 1$
string theories. In this case we can consider a disc with the holes and set
the boundary condition $e^{\varphi }\rightarrow \infty $ , recently
considered in [4]. According to this work the open strings with this
boundary conditions do not have excited states, at least for the simplest
bootstrap solution. To obtain the sum of planar diagrams we have to supply
these strings with the Chan- Paton factors. Hence, from the open string
point of view all degrees of freedom are described by the matrix model which
can be shown to coincide with the Kontsevich model. From the closed string
point of view we have a Liouville theory in a certain background. Along
these lines it is possible to resolve a long standing question - why the
matrix models describe the Liouville theory.

The picture we are describing is just the standard D-brane picture of gauge
fields (see e.g [3]). What we are trying to explain is why the stack of the
D-branes in flat space can be replaced by the closed string backgrounds. The
full quantitative derivation of this is still lacking, but we will see how
the above arguments work in the different situations. It is important to
realize though, that D-branes, being a beautiful and useful tool, are
neither necessary nor sufficient to derive even AdS/CFTcorrespondence, to
say nothing of the general case. A more powerful method to attack these
problems is based on the loop equations [5] But it also is not developed
enough to provide a complete understanding.

\section{The entropy of gauge invariant words.}

In this section we discuss the structure and combinatorics of the gauge
invariant words in preparation to the comparison with the string theory. A
simplest candidates for the gauge invariant words are expressions of the
type of $Tr(\nabla ^{k_{1}}F...\nabla ^{k_{n}}F)$where $F$ stands for the
Yang- Mills field strength and $\nabla $ for a covariant derivative. These
words are related to the closed string states. But this particular alphabet,
consisting of the letters $\nabla ^{k}F$ is not convenient for our purposes
because the letters are not independent. There are kinematical relations
coming from the Bianchi identities and from the relation $F_{\mu \nu
}=[\nabla _{\mu },\nabla _{\nu }]$. There are also the dynamical relations
coming from the Yang-Mills equations. A better alphabet, which will allow us
to count the words, is obtain by taking a radial gauge
\begin{equation}
x_{\mu }A_{\mu }=0
\end{equation}
By differentiating this relation and setting $x=0$ we find a set of letters
\begin{equation}
\Omega _{\alpha _{1}...\alpha _{n};\beta }=\partial _{\alpha
_{1}}...\partial _{\alpha _{n}}A_{\beta }
\end{equation}
with a single relation
\begin{equation}
\Omega _{\alpha _{1}...\alpha _{n};\beta }+permutations=0
\end{equation}
Therefore $\Omega $ transforms as (n,1) Young tableau. Let us consider first
the gauge theory without matter in the weak coupling limit and count the
number of letters of given dimension
\begin{equation}
\#(\Omega )=\sum_{\{n_{j}\}}(d\delta _{n,\sum n_{j}}-\delta _{n+1,\sum
n_{j}})
\end{equation}
Here $d$ is the number of dimensions, $n_{j}$ is the number of indices equal
to $j$ and the second term is the number of the constraints (9 ) which
reduce the number of $\Omega $ -s. To find the generating function for $%
\Omega $ we simply notice that in the weak coupling limit it has a dimension
$\Delta =n+1$ and thus contributes $x^{n+1}$. Therefore the partition
function $F(x)$ for the letters has the form
\begin{equation}
F(x)=\frac{dx}{(1-x)^{d}}-\frac{1}{(1-x)^{d}}+1
\end{equation}
If the matter fields are present, they can be easily included in this
counting. Now we have to find the partition function for the words. A slight
complication here is that the cyclic symmetry of the words must be properly
accounted. For example, a word $Tr(AABB)$ has four cyclic reincarnations and
thus must have a weight $\frac{1}{4}$ . At the same time a similar word $%
Tr(ABAB)$has only two cyclic versions and thus the weight is $\frac{1}{2}.$
This combinatorial problem can be handled by the Polya theory. It is
based on the observation that each transformation of the symmetry group
splits the set of words onto cycles of different length and introduces the
cyclic index
\begin{equation}
\Phi (z_{1}...z_{n}...)=\frac{1}{[G]}\sum_{g\in
G}z_{1}^{b_{1}}...z_{k}^{b_{k}}...
\end{equation}
where $G$ is the symmetry group (the cyclic permutations in our case), $[G]$%
is the number of its elements and $b_{k}$ is the number of the cycles of the
length $k$ . According to Polya, the partition function is expressed in
terms of the cyclic index by the formula
\begin{equation}
Z=\Phi (\sum x_{k},\sum x_{k}^{2},...\sum x_{k}^{n}...)
\end{equation}
Incidentally, the meaning of these formulas can be easily understood by the
use of path integrals with the sums over permutations,as was done by Feynman
to derive the Bose statistics.

Using these methods we obtain the following answer for the partition
function of the words
\begin{equation}
Z=-\sum_{n=1}^{\infty }\frac{\varphi (n)}{n}\log
(1-F(x^{n}))=-\sum_{n=1}^{\infty }\frac{\varphi (n)}{n}\log [\frac{1-dx^{n}}{%
(1-x^{n})^{d}}]
\end{equation}
Here $\varphi (n)$is the Euler function equal to the number of integers
relatively prime to $n.$ It takes care of the subtlety with the weights
described above.

In the case of the N=4 Yang Mills theory we have to add the scalar fields $%
\Phi _{i}$ to our alphabet the letters $\partial _{\alpha _{1}}...\partial
_{\alpha _{n}}\Phi _{i}$ where $i=1...10-d.$ An obvious modification of the
above counting gives in this case
\begin{equation}
Z=-\sum_{n=1}^{\infty }\frac{\varphi (n)}{n}\log [\frac{1-10x^{n}}{%
(1-x^{n})^{d}}]
\end{equation}
We do not include letters containing fermions, although it is also easy to
do. These formulas count the off-shell words not constrained by the equation
of motion, which generate some further relations between the words. To find
the on-shell entropy we notice that from the linearized Yang-Mills equations
\begin{equation}
\partial ^{2}A_{\alpha }-\partial _{\alpha }\partial _{\lambda }A_{\lambda
}=0
\end{equation}
it follows after multiple differentiation that
\begin{equation}
\Omega _{\beta _{1}...\beta _{n-2}\lambda \lambda ;\alpha }-\Omega _{\beta
_{1}...\beta _{n-2}\lambda \alpha ;\lambda }=0
\end{equation}
The number of extra constraints is now easy to count. The tensor in ( 17)
has zero trace when $\alpha $ and $\beta $ are contracted and otherwise all
its components are independent. By the same method as before we obtain the
partition function for the letters on shell, $\widetilde{F(x)}$%
\begin{equation}
\widetilde{F}(x)=F(x)-\frac{dx^{3}}{(1-x)^{d}}+\frac{x^{4}}{(1-x)^{d}}=1-%
\frac{(1-x^{2})(1-dx+x^{2})}{(1-x)^{d}}
\end{equation}
It is interesting to notice that for d=4 the on-shell $\Omega $ -s transform
according to ($\frac{n+1}{2},\frac{n-1}{2})\oplus (\frac{n-1}{2},\frac{n+1}{2%
})$ representation of SO(4). Correspondingly $\widetilde{F=}2\sum
n(n+2)x^{n+1}.$

Using these formulas we can evaluate the number of words ${\cal N}$ of the
given dimension $\Delta $ by the formula
\begin{equation}
{\cal N=}\oint \frac{dx}{x^{\Delta +1}}Z(x)
\end{equation}
When $\Delta $ is large the asymptotic is determined by the singularities of
$Z$ and we get (on shell)
\begin{eqnarray}
{\cal N} &\sim &e^{b\Delta } \\
b &=&\log (\frac{d}{2}-\sqrt{\frac{d^{2}}{4}-1})
\end{eqnarray}
An exponential behavior in a similar problem has been numerically observed
in [6]. This formula gives the nuber of states in the limit of zero gauge
coupling. In the opposite limit of the large gauge coupling $\lambda $ the
states are described by the almost free string theory. According to [7,8] in
this case $\Delta \sim \lambda ^{\frac{1}{4}}M,$ where $M$ is the mass of
the string state. Since the number of states in the closed string grows
exponentially with the mass, it is safe to conjecture that ${\cal N\sim }%
e^{b(\lambda )\Delta }$ with $b(\lambda )\rightarrow const$ at small $%
\lambda $ and $b(\lambda )\sim \lambda ^{-\frac{1}{4}}$ at large $\lambda .$

\section{Words and vertices}

The essence of the gauge fields - strings correspondence is the statement
that there is an isomorphism between the gauge invariant operators and the
vertex operators of a certain closed string theory in the background (6 ).
This statement is presumed to be true for a general gauge theory, although
it has been thoroughly checked only for the Yang- Mills theory with maximal
supersymmetry and in a few related cases. Even in these cases the derivation
from the first principles is still lacking. To set the stage let us review
this isomorphism, taking as an example conformal gauge theories, for which
the background metric is known [9 ]. Namely the conformal group on the gauge
side must be present as a group of motion for the metric on the string side.
That leads us to the space of constant negative curvature, the AdS. It is
convenient in this case to rewrite ( 6) in the form
\begin{equation}
ds^{2}=\sqrt{\lambda }\frac{dy^{2}+d\overrightarrow{x}}{y^{2}}^{2}
\end{equation}
where the Liouville field $\varphi \sim \log y$ and $\lambda $ measure the
curvature and will be related to the Yang-Mills coupling. For the case of
the four -dimensional gauge theories the symmetry group of this metric is
O(5.1). The string theory in question is described by the two-dimensional $%
\sigma $ -model on the world sheet the basic field of which is $n(\xi )$,
the field of the Lorentzian unit vector, $n_{0}^{2}-\overrightarrow{n}$ $%
^{2} $=1. The six dimensional unit vector $n$ is related to $x$ and $y$ in (
22) by the standard parametrization
\begin{eqnarray}
n_{-} &=&n_{0}-n_{1}=y^{-1} \\
n_{+} &=&n_{0}+n_{1}=y^{-1}(y^{2}+\overrightarrow{x}^{2}) \\
n_{\mu } &=&y^{-1}x_{\mu }
\end{eqnarray}
The Lagrangian of the $\sigma $ -model consists of several pieces. For the
case of the N=4 Y.-M. theory it contains, apart from the $n$ field
describing AdS$_{5}$ another field , $N,$describing six scalars and having
the $R$ symmetry O(6). This is a field of $S_{5}$ [ 9] , making the
background ten dimensional, AdS$_{5}\times $S$_{5}.$In the pure gauge theory
this field would be absent. However strings in the critical dimension (ten)
are simpler than in the non-critical one and this is precisely why presently
we know more about N=4 Yang -Mills theory then about pure gauge theory ( but
it is the latter that is our final goal ).We have the following structure
\begin{equation}
L=\frac{\sqrt{\lambda }}{2}[(\partial _{\alpha }n)^{2}+(\partial _{\alpha
}N)^{2}]+L_{F}(n)+L_{F}(N)+e^{-\frac{\chi }{2}}\sum (n)\sum (N)+L_{ghost}
\end{equation}
where $L_{F}$ is the standard Lagrangian for the NSR fermions which are
orthogonal to $n$ and $N$ correspondingly. The $L_{ghost}$ describes the
bosonic ghosts of the NSR string and the field $e^{-\frac{\chi }{2}}$ is the
spin operator for these ghosts, while the operators $\Sigma $ are the spin
operators for the NSR fermions. Notice that the fields $n$ and $N$ interact
only via the spin term (which creates antiperiodic boundary conditions for
all NSR fermions simultaneously). Without this interaction the sigma model
would not be conformally invariant on the world sheet. Instead the coupling
for the $N$ field would be asymptotically free, while the coupling for $n$
would flow in the opposite direction. The spin-spin interaction locks them
together.

The vertex operators are simply various primary operators of this CFT with
dimension (1,1). Let us discuss their structure, concentrating on the most
non-trivial ones formed out of the non-compact $n$ -field with the group
O(5,1). A convenient trick here is to start with the compact case of O(6)
and then perform an analytic continuation. The simplest operators are those
which do not contain derivatives (we will call them ''level zero''
operators). They have the form
\begin{equation}
V=\Psi _{i_{1}...i_{l}}n_{i_{1}}...n_{i_{l}}+...
\end{equation}
here $\Psi $ is a totally symmetric traceless tensor and we dropped the
fermionic terms which can mix (in higher orders) with the written one. The
anomalous world sheet dimension $\delta $ of this term in the one loop
approximation is given by the well known formula
\begin{equation}
\delta =\frac{1}{\sqrt{\lambda }}l(l+4)
\end{equation}
where the first factor is simply the sigma model coupling constant, while
the second is the Casimir operator for our representation. To perform the
analytic continuation let us consider the highest weight component in $V$%
\begin{equation}
V_{+}=n_{+}^{l}=y^{-l}(y^{2}+\overrightarrow{x}^{2})^{l}
\end{equation}
and notice that after going to O(5,1) the formula for the anomalous
dimension ( 28) changes sign, because we change the sign of the curvature;
so now
\begin{equation}
\delta =-\frac{1}{\sqrt{\lambda }}l(l+4)
\end{equation}
Another consequence of the non-compactness is that the angular momentum $l$
doesn't have to be an integer anymore. Moreover from (29 ) we see its new
meaning - it defines the space-time scaling of $V$ ( $V_{+}\sim x^{l}).$
However so far we found only the highest weigh component of the
representation. To find all other operators is very easy. We have to exploit
the fact that translations in $x$ (because they are the part of SO(5.1))
leave us inside the representation, and hence $V_{+}(x+a),$ where $a$ is an
arbitrary vector, transform under the same representation as $V_{+}(x).$ In
fact these objects with different $a$ form a complete basis for the
representation. It is convenient to make a linear transformation of this
basis by taking the Fourier transformation in $a$ and deal with the
operators
\begin{equation}
V_{p\Delta }=\int d^{4}an_{+}^{-\Delta }(x+a,y)e^{ipa}=y^{\Delta
}e^{ipx}\int d^{4}ae^{ipa}\frac{1}{(a^{2}+y^{2})^{\Delta }}=N(\Delta
)p^{\Delta -2}y^{2}K_{\Delta -2}(py)e^{ipx}
\end{equation}
where we introduced the space-time scaling dimension $\Delta =-l$ , $%
N(\Delta )$ is an irrelevant normalization factor, $K$ is the modified
Bessel function; we wrote these formulas for the four dimensional case,
while in the dimension d we would get $K_{\Delta -\frac{d}{2}}$ in them.
Under the O(5,1) transformations the operators with different $\Delta $ do
not mix, while $p$ plays the role of the magnetic quantum number. We see
from ( 30) that the space- time and the world sheet dimensions are related
as
\begin{equation}
\delta =\frac{1}{\sqrt{\lambda }}\Delta (4-\Delta )
\end{equation}
So far we treated the operators without world sheet derivatives (level
zero). Let us consider next level two operators. They have a general form
\begin{equation}
V=\Psi _{ab,i_{1}...i_{l}}n_{i_{1}}...n_{i_{l}}\partial _{z}n_{a}\partial _{%
\overline{z}}n_{b}
\end{equation}
These operators are reducible from the point of view of O(6), and transform
as a product of (1,0,0)$\times (1,0,0)\times (l,0,0)$ , where ( $%
f_{1}f_{2}f_{3})$is the representation with the lengths of the rows in the
Young tableau equal to these numbers (recall that O(6) has rank 3). A
convenient way to parametrize these numbers is to consider first the Lorentz
properties of $V.$ Since O(4)$\approx $ O(3)$\times $O(3), its
representations are labeled by the two integer angular momenta $%
(j_{1},j_{2}) $. In order to characterize a representation of O(5,1) we
simply add $\Delta $ to this list, getting a triple $(j_{1},j_{2,}\Delta ).$
The world sheet dimensions are in general different for different
representations, although in the limits of strong and weak couplings certain
degeneracies occur. For the operators ( 33) we have the formula
\begin{equation}
\delta =1+\frac{1}{\sqrt{\lambda }}[\Delta (4-\Delta )+\varepsilon
(j_{1},j_{2})]+o(\frac{1}{\sqrt{\lambda }})
\end{equation}
In this formula the first term is just the naive dimension of ( 33), the
second term , as in ( 30) , represents the center of mass motion of the
string on AdS$_{5}$ and the correction $\varepsilon (j_{1},j_{2})$ (which is
not hard to calculate but it has not been done) is the interaction between
the center of mass motion of the string and its oscillations. Notice that
this last effect is not present in the flat space where the Hamiltonian is
the sum of the oscillator part and the center of mass part. In the curved
space such a separation does not exist and $\varepsilon $ -like corrections
are different for different levels. We will further discuss it in section 5.

By the same method as before we can express our operator in terms of $%
\overrightarrow{x}$ and $y$ . Let notice that in the level zero case we can
describe the operator $V_{p\Delta }(\overrightarrow{x}$ $,y)$ as a an
eigenfunction of the Laplace operator with the asymptotic behaviour $%
V_{p\Delta }\longrightarrow y^{-\Delta }$ as $y\rightarrow 0$. The level two
operator has the structure
\begin{equation}
V\sim y^{-\Delta -2}\Psi _{mn}(py)\partial _{z}x^{m}\partial _{\overline{z}%
}x^{n}e^{ipx}
\end{equation}
where $m,n=0,...4$ and we define $x^{0}=y$ ; we also adjusted the power of $%
y $ so that the space- time dimension of $V$ is equal to $\Delta .$ As we
explained above, for a fixed $\Delta $ the operator must have definite
Lorentz properties . For a tensor operator $V_{p\Delta }^{\mu \nu }$we must
choose the function $\Psi _{mn}^{\mu \nu }$ in such a way that it is an
eigenfunction of the tensor Laplacian in AdS and has the asymptotic
behaviour at $y\rightarrow 0$
\begin{equation}
\Psi _{\lambda \gamma }^{\mu \nu }\rightarrow \delta _{\lambda }^{\mu
}\delta _{\gamma }^{\nu }
\end{equation}
Notice that (36 ) specifies only spatial components of $\Psi $ The gauge
conditions satisfied by $V$ will determine the ''time'' components which
can't be fixed independently. As a result we obtain a uniquely defined
tensor vertex operator. In general we should expect that the traceless part
of $V^{\mu \nu }$ and its trace have different dimensions. However in the
special case of N=4 supersymmetry they coincide, being related to the chiral
operators.

Continuing in this way we obtain all possible operators of the above sigma
model. The vertex operator form a subset defined by the conditions that they
are conformal primaries on the world sheet and have dimensions (1,1). The
last condition gives us an equation for determining their space-time
dimensions
\begin{equation}
\delta (j_{1},j_{2},\Delta )=1
\end{equation}
This is the on-shell condition. It is important to realize that the 4d
momentum $p$ is not constrained by this condition. Sometimes it is more
convenient to use the coordinate representation for the integrated vertex
operator which (omitting the Lorentz indices) has the structure
\begin{equation}
\Omega (\overrightarrow{x})=\int d^{2}\xi V(\overrightarrow{x}+%
\overrightarrow{x}(\xi ),y(\xi ))
\end{equation}
The gauge - strings duality is the isomorphism between the set of operators $%
\Omega $, obtained from the various (1,1) primaries of the above sigma model
and the gauge- invariant words. Let us explain the above conjectures, using
(to shorten the notations) the ''holomorphic halves'' of the vertex
operators. On the level one we consider an operator
\begin{equation}
\Omega _{1}=\Psi _{m}(x)\partial _{z}x^{m}
\end{equation}
For small curvatures we can use the standard technic [ 10] to determine the
change $\stackrel{.}{\Omega }$ of this operator when we apply an external
Liouville field $\sigma .$ A simple one loop computation gives
\begin{equation}
\stackrel{.}{\Omega }=\sigma \nabla ^{2}\Psi _{m}\partial _{z}x^{m}+\partial
_{z}\sigma \nabla ^{m}\Psi _{m}
\end{equation}
The first term comes from the action of $L_{0}$ Virasoro generator while the
second - from $L_{1}.$ If we require that the operator should be $\sigma $ -
independent, that is conformal primary, we get the equations
\begin{eqnarray}
\nabla ^{2}\Psi _{m} &=&0 \\
\nabla ^{m}\Psi _{m} &=&0
\end{eqnarray}
We see that the second equation determines the $y$ -component of $\Psi $ ,
while the solution of the first one is determined by the asymptotic
condition
\begin{equation}
\Psi _{\mu }\sim y^{\Delta -1}\psi _{\mu }(x)
\end{equation}
The exponent $\Delta -1$ (the value of which is determined by the on-shell
condition (37) ) is defined so that $\psi _{\mu }(x)$ scales like $%
x^{-\Delta }$ . This follows from the fact that $\Psi _{m}$ , being a vector
in AdS, scales as $x^{-1}$ under its isometries. Let us also notice that if
we require the $\sigma $ -independence only of the integrated $\Omega $ we
get a gauge unfixed version of the equations ( 41, 42).

The full vertex operator is defined by the function in the 4d space $\psi
_{\mu }(x)$or by its Fourier transform $\psi _{\mu }(p).$ A corresponding
object in the standard string theory would be a photon polarization. It is
important to realize that while any $\psi _{\mu }$ generates a physical
vertex operator, some of these operators can have zero norm. In the above
simple example this happens in the following way. The above vertex has the
form
\begin{equation}
V_{\mu }(p)=\int dz\Psi _{\mu m}(p,y)\partial _{z}x^{m}e^{ipx}
\end{equation}
where the function $\Psi _{m\mu }(p,y)$is a solution of (41 ) with the
asymptotic conditions
\begin{equation}
\Psi _{\lambda \mu }(p,y)\rightarrow \delta _{\lambda \mu }y^{\Delta -1}
\end{equation}
From this we conclude that
\begin{equation}
p_{\mu }\Psi _{\mu m}(p,y)=\nabla _{m}\chi (p,y)
\end{equation}
if $\chi $ satisfies the Laplace equation with the conditions $\chi
\rightarrow y^{\Delta -1}.$ After plugging ( 46) into ( 44) we see that the
vertex operator satisfies the conservation equation
\begin{equation}
p_{\mu }V_{\mu }=0
\end{equation}
This is just a rephrasing of the statement that in the critical string
theory we have a zero norm state at the level one, generated by the Virasoro
operator $L_{-1}$ . To apply it to our closed string case we have to do the
usual doubling of the vertex $V_{\mu }$ and obtain the tensor vertex (33 ).
Then the vertex is identified with the energy momentum tensor of the gauge
theory and the existence of the null state on the string side implies a
conservation law on the gauge theory side. As we proceed to the level two,
we expect that another null state, generated by the $L_{-2}+\frac{3}{2}%
L_{-1}^{2}$ will play the similar role, providing some relations between
gauge- invariant words. The (holomorphic half) at the level two has the form
\begin{equation}
\Omega =\psi _{m}\nabla _{z}\partial _{z}x^{m}+\chi _{mn}\partial
_{z}x^{m}\partial _{z}x^{n}
\end{equation}
In the one loop approximation (small AdS curvature) it is easy to find the
physical state conditions. In the expansion similar to ( 40) we find that
the change of $\Omega $ with the respect to the external Liouville field has
the form
\begin{equation}
\stackrel{.}{\Omega }=\sigma L_{0}\Omega +\partial _{z}\sigma L_{1}\Omega
+\partial _{z}^{2}\sigma L_{2}\Omega +(\partial _{z}\sigma
)^{2}L_{1}^{2}\Omega
\end{equation}
This formula can be easily derived by passing to the light-cone gauge on the
world sheet (I didn't find it in the literature). To ensure that the state
is physical we have to calculate (using the method of [ 10] ) the
corresponding terms and set them to zero. That gives
\begin{eqnarray}
(\frac{1}{\sqrt{\lambda }}\nabla ^{2}-1)\psi _{m} &=&(\frac{1}{\sqrt{\lambda
}}\nabla ^{2}-1)\chi _{mn}=0 \\
\nabla ^{m}\psi _{m}+\frac{1}{2}\chi _{k}^{k} &=&0 \\
\frac{1}{\sqrt{\lambda }}\nabla ^{m}\chi _{mn}+\psi _{m} &=&0
\end{eqnarray}
Once again, the solution of these equations is specified by fixing the
asymptotic behaviour of the 4d components of these tensor fields, $\psi
_{\mu }(\overrightarrow{x}$) and $\chi _{\mu \nu }(\overrightarrow{x}$).
Similarly to the scalar case (38) we have the vertex operators $V_{\mu }(%
\overrightarrow{x})$ and $V_{\mu \nu }(\overrightarrow{x})$ given by the
formulas
\begin{equation}
V_{\mu }(\overrightarrow{x})=\int dz(g_{\mu m}(\overrightarrow{x}+%
\overrightarrow{x}(z),y(z))\partial _{z}x^{m}+h_{\mu mn}(\overrightarrow{x}+%
\overrightarrow{x}(z),y(z))\partial _{z}x^{m}\partial _{z}x^{n})
\end{equation}
and analogously for $V_{\mu \nu }.$ Here $g$ and $h$ are the corresponding
propagators for the equations ( 50 - 52).

The existence of the null vectors at the levels one and two implies that
certain linear combinations of these vertex operators have zero matrix
elements. It is easy to see that while the above vertices from the point of
view of SO(4) transform as $(\frac{1}{2},\frac{1}{2})\oplus (1,1)\oplus
(0,0),$ the elimination of the null states leaves only $(1,1)$
representation. This is because the corresponding gauge parameters are
scalar (at the level two) and vector (at the level one).

Our main conjecture is that any physical ( that is satisfying the Virasoro
constraints) vertex operator corresponds to a certain combination of the
gauge-invariant words. Setting the null states to zero generates certain
linear combinations between the vertex operators. These relations must be
equivalent to the Yang-Mills equations of motion. In other words the
condition
\begin{equation}
<V_{null}...>=0
\end{equation}
can be considered as a variational principle for the D-branes describing the
Yang-Mills theory.

To clarify this statement let us consider as an example the null state
generated by $L_{-1}$ and assume that we have a D-brane located at $Y^{i}=0$
, where $Y$ is a transverse coordinate. It is well known that the above null
state corresponds (in the closed strings) to an infinitesimal
diffeomorphisms $Y^{i}\Rightarrow Y^{i}+\varepsilon ^{i}(x)$ in the sense
that its insertion in the middle of the world sheet disc describing the
D-brane generates this transformation at the boundary. Let $S_{D}[Y]$ be the
Born-Infeld action describing the D-brane. Then the condition that the null
state decouples from the D-brane is that the action remains unchanged under
this diffeomorphism, or $\frac{\delta S_{D}}{\delta Y^{i}}=0.$

The check of our conjecture requires a better understanding of the
correspondence between the words and the string states then we presently
have. The problem is to relate the words in the weak Yang- Mills coupling
limit, described by the formulae (8) and (17), to the string states, which
are easy to describe in the opposite limit. Only after that we will be able
to translate explicitly the condition ( 54) to the gauge theory language.

While the full correspondence is unknown, we can formulate a principle which
governs it. Consider the (space-time) anomalous dimensions $\Delta
_{1}(\lambda )$and $\Delta _{2}(\lambda )$of two operators $\Omega _{1}$and $%
\Omega _{2}$, where once again $\lambda $ is the gauge coupling. As we
change $\lambda $ from zero to infinity these dimensions flow from the fixed
integer values, defined by the free fields to the arbitrary large values
following from the formula ( 37) (we consider unprotected operators). We
shall use the following '' {\it non-intersection principle}'' : the
trajectories of the operators with the same symmetry do not intersect. It is
easy, by the use of the conformal perturbation theory, to reduce this
statement to the standard secular equation argument of quantum mechanics. As
an example of an application of this principle , consider (modulo fermions
and gauge fields) an operator $Tr(F_{\alpha \beta }F_{\gamma \delta })$ from
the point of view of SO(4) its content is given by the representations $%
2(0,0)\oplus (1,1)\oplus (2,0)\oplus (0,2).$ Scalar and tensor
representations correspond to the graviton, dilaton and axion and are
protected by the supersymmetry. The unprotected operator is thus $%
(2,0)\oplus (0,2)$. Now as the coupling increases it must go to the first
available level of the weakly coupled string theory. This is because
otherwise this level will be taken by an operator with the same SO(4)
quantum numbers but with the higher naive dimension. In these circumstances
the intersection of the trajectories $\Delta _{1}(\lambda )$ and $\Delta
_{2}(\lambda )$ will be unavoidable. As we saw, at the level two we have a
state $(1,1)$ in the holomorphic part of the vertex operator. When we take a
product of the holomorphic and antiholomorphic parts we get among other
representations the desired $(2,0)\oplus (0,2).$ We conclude that it must be
occupied by our gauge operator.

However, in order to check quantitatively that at each level decoupling of
the null states corresponds to passing from the off -shell gauge invariant
words to the on-shell ones, one needs a much finer combinatorial analyses.
Namely, it is necessary to find the number of the gauge operators of given
dimension and given SO(4) quantum numbers and compare them with the
corresponding decomposition on the string side. This work is now in progress.

\section{Weak and strong coupling}

The gauge / string correspondence is a duality in the sense that the strong
coupling limit on the gauge side corresponds to the weak coupling in string
sigma model (that is a small curvature of the target space) and vice versa.
The main difficulty of the subject stems from our temporary inability to
solve exactly the above sigma model and to analyze gauge theory in terms of
its physical states. In this section we will introduce some further
conjectures concerning these states.

Let us notice first of all, that the strong coupling limit must be simple.
Indeed, the spectrum of the space- time anomalous dimension in this case
must coincide with that of the free gauge fields and thus all the dimensions
must be integer. At the same time, formulae (32) and (37) show that at large
$\lambda $ we have a behaviour $\Delta \sim \lambda ^{\frac{1}{4}}$ for a
very large $\lambda $ [ 7,8] corresponding to the small curvature limit (the
space-time curvature $R\sim \frac{1}{\sqrt{\lambda }}$ ). If we try to
extrapolate blindly this formula to the small $\lambda $ we would get a
completely wrong behaviour - namely all space -time dimensions will become
degenerate. This happens because in this extrapolation the world-sheet
dimensions blow up in this limit.

Another strange feature of the above result is the following. Consider large
but finite $\lambda $ in gauge theory. It seems that typically in
perturbation theory for the operators of very high dimension and twist the
normal (integer) part of dimension is larger then the anomalous
contribution. Hence it is natural to expect that at any $\lambda $ we have
highly excited states retaining (asymptotically) their normal dimensions.

We conjecture here that the resolution of these puzzles lies in the
structure of the spectrum of anomalous dimensions in the sigma model.
Namely, we assume that while the dimension of a generic operator blows up in
the strong coupling limit, there is a special subset of operators with
finite dimensions. These operators have integer space-time dimensions and
correspond to the free gauge fields. The same operators dominate at finite $%
\lambda $ but at very high levels.

Let us discuss a possible origin of these special operators. As a first
example consider an operator of the form
\begin{equation}
V=n_{+}^{-\Delta }(\partial _{z}n\partial _{\overline{z}}n)^{s}(\partial
_{z}n)^{2p}(\partial _{\overline{z}}n)^{2p}+...
\end{equation}
where the dots refer to the similar terms with smaller $s$ but the same
overall dimension. In the one loop approximation dimensions of these
operators have a peculiar feature first noticed in the context of
localization theory in [ 11] and in O(d) sigma models in [12 ]. Adapting the
results of these papers to our case we get the following formula for
anomalous dimensions
\begin{equation}
\delta =s+2p+\frac{1}{\sqrt{\lambda }}\left[ \Delta (4-\Delta
)+2s(s-1)-16p\right] +o(\frac{1}{\sqrt{\lambda }})
\end{equation}
We see that at large $s$ we have from $\delta =1$ condition (if we trust the
one loop approximation) $\Delta \approx \sqrt{2}s$ independently of $\lambda
.$ We also see that due to the peculiar sign of the $s$ -dependent term,
which tends to decrease the dimension in the compact case, the world sheet
dimension $\delta $ doesn't blow up as $\lambda \rightarrow 0.$ There are
other operators with the same properties and different Lorentz structure.

In the one loop approximation world sheet supersymmetry and RR backgrounds
don't change these results. However we can consider them only as a hint of
the true structure, before we have higher approximation under control. This
problem is not solved yet. In what follows I shall describe a possible
approach to its solution.

If we set $\lambda $ to zero in the eq. (26) we are left with the fermionic $%
\lambda $ -independent Lagrangian.. Hopefully it will describe the above
subset of operators with finite dimensions. The structure of the $L_{F}$
term is as following
\begin{equation}
L_{F}=\overline{\psi }\gamma _{\mu }(\partial _{\mu }+\omega _{\mu }+A_{\mu
})\psi +\sqrt{\lambda }A_{\mu }^{2}
\end{equation}
where $\omega _{\mu }$ is a spin-connection on the hyperboloid in question
and $A_{\mu }$ is an auxiliary field needed to reproduce the four fermion
term in the supersymmetric sigma model. We see that as $\lambda \rightarrow
0 $ the last term disappears and we can shift $A_{\mu }+\omega _{\mu
}\Rightarrow A_{\mu }$. As a result we end up with the fermions with ''zero
current'' condition, the system which is easy to analyze. This system, which
is equivalent to the Thirring model at infinite coupling, in which only
gauge invariant operators have finite dimensions and they can be classified.
Unfortunately this doesn't solve our problem. We have to treat the spin-spin
(or RR) interaction in (26). The zero current systems with the RR
-interaction has not been studied so far, although they are much simpler
then the original sigma model. The spin operators should play a very
important role in this analyses, since their dimensions, being related to
the central charge, do not blow up. Perhaps they form the building blocks
for the free gauge fields.

Finally, let us stress that although in this section we dealt with the
theories which are conformally invariant in space-time and seemingly
excluded the most interesting cases in which this invariance is broken, this
is not a serious limitation. Because of the asymptotic freedom all theories
are almost invariant in the large curvature limit. More technically one can
consider the $4+\varepsilon $ dimensional space in which the above symmetry
is strict.

\section{The Outlook}

The picture which slowly arises from the above considerations is that of the
space-time gradually disappearing in the regions of large curvature. The
natural description in this case is provided by a gauge theory in which the
basic objects are the texts formed from the gauge-invariant words. The
theory provides us with the expectation values assigned to the various
texts, words and sentences. These expectation values can be calculated
either from the gauge theory or from the strongly coupled 2d sigma model.
The coupling in this model is proportional to the target space curvature.
This target space can be interpreted as a usual continuous space-time only
when the curvature is small. As we increase the coupling, this
interpretation becomes more and more fuzzy and finally completely
meaningless. Since the theory is not complete, we can't give an explicit
demonstration of this mechanism.

Apart from the cases considered above it is perhaps worthwhile to examine an
old example of the similar phenomenon given in [13 ]. In this work I have
looked at the sigma model with the Lagrangian
\begin{equation}
L=(\partial \varphi )^{2}+a^{2}(\varphi )(\partial \overrightarrow{N}%
)^{2}+...
\end{equation}
This theory has two interpretations. First, in the weak coupling domain it
describes the Friedman universe with positive curvature. The one loop
approximation for the function $a$ gives the Einstein-like equations for
this metric and not surprisingly the solution has a Big Bang singularity at
some $\varphi =\varphi _{0}$ , $a(\varphi _{0})=0.$ In the second
interpretation this is an O(3) sigma model coupled to 2d gravity with $%
a^{-2} $ being a running coupling constant. The one loop renormalization
group once again gives a singularity due to the asymptotic freedom. But this
singularity is a fiction! As well known, the $\overrightarrow{N}$ field
develops a mass gap and the theory is completely non-singular . However this
target space can not be interpreted as a 4d space-time. Instead it is
characterized by a set of amplitudes derived from the above sigma model,
which has the space-time interpretation only in the quasi-classical domain.
In the very early universe space-time is not getting singular - it simply
doesn't exist. It would be very interesting to find the gauge theory
corresponding to the above model. The running of $a(\varphi )$is related to
the running of the gauge coupling.

Another related problem is to give the gauge theory interpretation to the
conjectured [ 13] infrared screening of the cosmological constant. This
screening must also be related to the gauge coupling, since this quantity
measures the curvature of space-time. Unfortunately we are still far from
finding the ultimate gauge theory which describes the Universe.

I am grateful to B. Altshuler for drawing my attention to the papers
[11,12]. This work was supported by the NSF grant PHY9802484.
\newpage
REFERENCES

[1] K. G. Wilson Phys. Rev. D10, 2445 (1974)

[2] A. M. Polyakov Int. Journ. of Mod. Phys. A 14 (1999) 645 [
hep-th/9809057]

[3] O.Aharony, S. Gubser, J. Maldacena, H. Ooguri , Y. Oz Phys. Rept. 323,
183 (2000)

[4] A. Zamolodchikov, Al. Zamolodchikov hep-th/0101152

[5] A.\ M. Polyakov, V. Rychkov Nucl. Phys.B581 (2000) 116 [ hep-th/0002106]

[6] D. Gross, Ig. Klebanov, A. Matytsin, A. Smilga Nucl. Phys. B461
(1996)109[ hep-th/9511104]

[7] S. Gubser, Ig. Klebanov, A.M. Polyakov Phys.Lett B428 (1998) 105
[hep-th/ 9802109]

[8] E. Witten Adv. Theor. Phys.2 (1998) 253 [hep-th/ 9802150]

[9] J. Maldacena Adv. Theor. Phys. 2 (1998) 231 [hep-th / 9711200]

[10] C. Callan, Z. Gan Nucl. Phys.B 272 (1986) 647

[11] V. Kravtsov, I. Lerner, V. Yudson Phys. Lett. A134 (1989) 245

[12] F. Wegner Z. Phys. B78 (1990) 33

[13] A. M. Polyakov Proceedings of Les Houches (1992) [hep-th /9304146]

\end{document}